# Genetic algorithm based optimization and post optimality analysis of multi-pass face milling

Sourabh K. Saha[*]
Department of Mechanical Engineering
IIT Kanpur
Kanpur, 208016, India

**Abstract:**
This paper presents an optimization technique for the multi-pass face milling process. Genetic algorithm (GA) is used to obtain the optimum cutting parameters by minimizing the unit production cost for a given amount of material removal. Cutting speed, feed and depth of cut for the finish and rough passes are the cutting parameters. An equal depth of cut for roughing passes has been considered. A lookup table containing the feasible combinations of depth of cut in finish and rough passes is generated so as to reduce the number of variables by one. The resulting mixed integer nonlinear optimization problem is solved in a single step using GA. The entire technique is demonstrated in a case study. Post optimality analysis of the example problem is done to develop a strategy for optimizing without running GA again for different values of total depth of cut.

# 1. Introduction

Optimization of cutting parameters in machining processes has been an important area of research. Starting from Taylor [1], there has been an increasing interest in machining process optimization [2-19]. Selection of optimum cutting conditions can lead to a substantial reduction in the operating costs. Commonly used objective functions are the minimization of production cost, the maximization of production rate and the maximization of profit rate. Abuelnaga and El-Dardiry [4] discussed a number of traditional optimization methods and highlighted the relative advantages and disadvantages of the methods for solving the problems of machining economics. These objectives can be represented in terms of the machining parameters such as cutting speed, feed, depth of cut and the number of passes. During selection of the parameters, care must be taken to ensure that the essential constraints are satisfied. Cutting force, power, surface finish and tool life are some of the commonly considered constraints [5-8]. Initial research in machining process optimization focused on single pass operations [1-4]. However, due to restrictions on the amount of material that can be removed in a single pass, multi-pass operations are required [6-10]. Machining is done in two stages in multi-pass operations. Majority of material removal takes place in a series of rough passes in the first stage. In the next stage a small amount of material is removed by a single finish pass. Typically, two distinct approaches have been used for optimization of multi-pass operations. One is the equal depth of strategy in which the depth of cut in all the rough passes is considered to be same [11, 12]. The other is the unequal depth of cut strategy. Typically such problems are optimized in two steps [13-15]. First, the optimum cutting speed and feed are obtained for all the possible depths of cut. Then, the total depth of cut is distributed optimally among the different rough and finish passes. Jawahir and Wang [16] have summarized the recent contributions to the field of machining process modeling and optimization. They have also presented a machining process optimization method in which many process performances such as surface roughness, cutting force, tool life and material removal rate have been combined into a single objective using weight factors.

*Corresponding Author email: sourabh.k.saha@gmail.com

Many researchers have dealt with the problem of machining parameters for turning process. Relatively less research has been done in optimization of multi-point cutting operations such as milling. Recently there have been quite a few reports in literature on optimization of milling operations [12, 14-19]. Sonmez et al. [12] used dynamic programming to obtain the optimum number of passes and geometric programming for obtaining the optimum cutting conditions. Their work showed that the performance of multi-pass milling operations is better than that of single pass operations. Shunmugam et al. [14] minimized the production cost for rough passes and finish passes in two stages. In the first stage, separate minimum costs for an individual rough and finish pass were determined and tabulated for various fixed values of depth of cut selected from a series of depths. In the second stage, genetic algorithm (GA) was used for finding the optimum number of rough passes and allocation of total stock in each of the rough passes and the final pass to achieve a minimum total production cost. An and Chen [15] used a similar technique for first stage and integer programming for the second stage for solving the same problem. The two stage strategy for optimization leads to a large number of computations and increases the computational time when there are a large number of passes to consider. Wang et al. [17] have presented a new methodology involving the use of GA for the selection of cutting conditions for multi-pass face milling operations based on a comprehensive criterion of integrating the contributing effects of all major machining performance measures. They first used Taguchi method for design of experiments to predict machining performance measures and then utilized genetic algorithms to optimize the cutting conditions. In their method, optimization of parameters in the rough and finish passes have been done simultaneously instead of the two stage procedure. Since Genetic Algorithms suffer from the problem of premature convergence to local optimum, some researchers have focused on using Hybrid Genetic Algorithms in which a local search based optimization procedure is used in conjunction with GA for optimizing machining operations [18, 19]. Wang et al. [19] used genetic simulated annealing (GSA) for multi-pass end milling operation to minimize the production time. GSA combines the properties of genetic algorithms (GA) with those of Simulated Annealing (SA) which is a local search based procedure. It was shown by them that the hybrid method produced results superior to those obtained by using GA alone.

In the present work, binary coded genetic algorithm (GA) with an elitist replacement strategy has been used for minimization of unit production cost in a multi-pass face milling operation. For a given amount of material removal (quantified by the total depth of cut), the optimum values for the cutting parameters have been obtained in a single step by using a lookup table method. The lookup table is generated by considering all the feasible combinations of depths of cut in the rough and finish passes for a given total depth of cut. All rough passes have been considered to be of equal depth of cut. Using GA, the cutting speed, feed and depth of cut in the finish and rough passes have been optimized simultaneously. The face milling operation and the cutting process model have been discussed in Section 2. In Section 3, the optimization technique based on GA has been explained. The technique has been demonstrated in a case study discussed in Section 4. Results obtained from the proposed technique have been compared with reported values of Shunmugam et al. [14] and An and Chen [15]. A numerical test has been performed to verify the convergence of the GA to the global optimum. Subsequently a constraint sensitivity analysis has been performed. This analysis is helpful in deciding on the relative importance of the constraints when the constraint limits are flexible. Through an analysis of the optimization results, an estimation strategy has been developed for quickly obtaining the optimum values without running GA, when the total depth of cut is changed. Although a multi-pass face milling process has been chosen as an example in this work, the lookup table based GA technique discussed here can be easily applied to other multi-pass machining processes.

## 2. Problem Formulation

### 2.1. Process Schematic

In face milling, the cutter is mounted on a spindle having an axis of rotation perpendicular to the work piece surface. The milled surface results from the action of cutting edges located on the periphery and face of the cutter.
In multi-pass milling the entire material is not removed in a single pass but in a series of passes which involves multiple rough passes and a final finishing pass.
The schematic representation of the cutting process is shown in Fig. 1.

The cutting parameters can be chosen to be different in each of the individual passes. However for simplification, all rough passes have been considered to be identical. Thus we have two sets of cutting parameters (speed, feed, and depth of cut), one for all the rough passes and one for the finish pass. Along with this, we have the number of rough passes as an additional variable. In this paper, for a given total depth of cut the unit production cost has been minimized to obtain the optimum values of the cutting parameters.

In rough machining, the length of the cutter travel is given by

$$L_{tr} = L + a_p + e_r \tag{1}$$

Where $L$ is the length of the work piece, $a_p$ is the approach and $e_{r,s}$ is the extra-travel of the cutter at the ends, normally taken to be 2–5 mm. The approach distance for symmetrical milling is given as

$$a_p = \frac{D}{2} - \sqrt{\left(\frac{D}{2}\right)^2 - \left(\frac{B}{2}\right)^2} \quad L_{ts} = L + D + e_s$$

(2)

In finish machining,

$$L_{ts} = L + D + e_s \tag{3}$$

as the cutter has to completely clear the work piece length [20].

## 2.2. Cutting Process Model

### 2.2.1. Decision Variables

Unit production cost depends on the cutting parameters used for machining. The cutting parameters which can be controlled are the cutting speed ($V_s$, $V_r$), feed ($f_r$, $f_s$) and depth of cut ($d_s$, $d_r$) for the finish and rough passes respectively and the number of rough passes (n). All these parameters are chosen as the decision variables.

### 2.2.2. Objective Function

The unit production cost (UC) can be represented as:

$$UC = CM + CI + CR + CT \tag{4}$$

Where
CM, CI, CR and CT are actual machining cost, machine idle cost, tool replacement cost and tool cost respectively [6]. These can be represented as:

$$CM = k_0(t_{ms} + nt_{mr}) \tag{5}$$

Where

$$n = \frac{(d_t - d_s)}{d_r}$$

$$t_{ms} = \frac{\pi D L_{ts}}{1000 V_s f_s Z}$$

$$t_{mr} = \frac{\pi D L_{tr}}{1000 V_r f_r Z}$$

(6)

$$CI = k_0 t_l$$
$$t_l = t_p + t_i$$
$$t_l = t_p + n(h_1 L_{tr} + h_2) + (h_1 L_{ts} + h_2)$$
$$CI = k_o[t_p + n(h_1 L_{tr} + h_2) + (h_1 L_{ts} + h_2)] \tag{7}$$

$$CR = k_o t_e Z \frac{t_{ms}}{T_s} + n(k_o t_e Z \frac{t_{mr}}{T_r}) \quad (8)$$

$$CT = k_t Z \frac{t_{ms}}{T_s} + n(k_t Z \frac{t_{mr}}{T_r}) \quad (9)$$

The tool life in finish and rough passes ($T_s$ and $T_r$ respectively) for face milling as discussed by Nefedov [20] can be expressed as:

$$T^l = \frac{C_v K_v D^{q_v}}{V d^{x_v} f^{y_v} B^{s_v} Z^{p_v}}$$

Where, $C_v$ and $K_v$ are constants, $D$ is the cutter diameter, $B$ is the width of cut, $Z$ is the number of teeth in the cutter; $l$, $p_v$, $q_v$, $s_v$, $x_v$, $y_v$ are constant exponents.

For a given cutter and fixed width of cut, the tool life for finish and rough conditions can be simplified in terms of the cutting parameters as:

$$T_s = \frac{C_o}{V_s^{n_1} d_s^{n_2} f_s^{n_3}}$$

$$T_r = \frac{C_o}{V_r^{n_1} d_r^{n_2} f_r^{n_3}} \quad (10)$$

Where the constant $C_o = \left(\frac{C_v K_v D^{q_v}}{B^{s_v} Z^{p_v}}\right)^{1/l}$ and, $n_1 = \frac{1}{l}, n_2 = \frac{x_v}{l}, n_3 = \frac{y_v}{l}$

Thus, the objective can be stated as
Minimize $UC = UC_s + nUC_r + k_o t_p$
Where

$$UC_s = \frac{a_s}{V_s f_s} + b_s V_s^{n_1-1} d_s^{n_2} f_s^{n_3-1} + c_s$$

$$UC_r = \frac{a_r}{V_r f_r} + b_r V_r^{n_1-1} d_r^{n_2} f_r^{n_3-1} + c_r \quad (11)$$

$a_s$, $b_s$, $c_s$, $a_r$, $b_r$ and $c_r$ are constants with units such that when 'V' is in m/min, 'f' in mm/tooth and 'd' in mm gives UC in $/piece.

### 2.2.3. Constraints

Variable Bounds

The parameters are allowed to vary in a limited range of values determined by the cutting tool manufacturer and the machine specifications.

- Cutting Velocity
$$V_{s,min} \leq V_s \leq V_{s,max}$$
$$V_{r,min} \leq V_r \leq V_{r,max}$$
$$\quad (12)$$

- Feed
$$f_{s,\min} \leq f_s \leq f_{s,\max}$$
$$f_{r,\min} \leq f_r \leq f_{r,\max}$$
(13)

- Depth of cut
$$d_{s,\min} \leq d_s \leq d_{s,\max} \text{ in steps of } d_{s,step}$$
$$d_{r,\min} \leq d_r \leq d_{r,\max} \text{ in steps of } d_{r,step}$$
(14)

Inequality constraints

- Force constraint

For milling process cutting force is given by
$$F = \frac{C_f K_f B^{s_f} Z^{p_f} d^{n_4} f^{n_5}}{D^{q_f}}$$

Where, $C_f$, $K_f$ are constants; $s_f$, $p_f$, $q_f$, $n_4$, $n_5$ are constant exponents. Representing the force in terms of only the machining conditions:

$$F = C_1 d^{n_4} f^{n_5} \quad (15)$$

Where the constant $C_1 = \dfrac{C_f K_f B^{s_f} Z^{p_f}}{D^{q_f}}$

Therefore,
$$F_s = C_1 d_s^{n_4} f_s^{n_5} \leq F_{\max}$$
$$F_r = C_1 d_r^{n_4} f_r^{n_5} \leq F_{\max}$$
(16)

Where $F_{\max}$ is obtained from machine tool limitations.

- Cutting power constraint
$$P = \frac{FV}{6120\eta} \leq P_{\max} \quad (17)$$

Neglecting contribution of feed force to power (due to low values of feed rate as compared to the cutting velocity) and using expression for force F from Eq. 15

$$P = C_2 V d^{n_4} f^{n_5}, \text{ where } C_2 = \frac{C_1}{6120\eta}$$

Hence,
$$C_2 V_s d_s^{n_4} f_s^{n_5} \leq P_{\max}$$
$$C_2 V_r d_r^{n_4} f_r^{n_5} \leq P_{\max}$$
(18)

- Surface finish constraint

Following Boothroyd [21], the surface finish for milling process can be expressed by:

$$R_a = 0.0321 \frac{f^2}{r_e} \tag{19}$$

And the surface requirement constraint is:

$R_a \leq R_{max}$

This can be reformulated in terms of feed as:

$$f_s \leq \sqrt{\frac{R_{s,max} r_e}{0.0321}}$$

$$f_r \leq \sqrt{\frac{R_{r,max} r_e}{0.0321}}$$

This constraint can therefore be merged with the variable bounds for feed as:

$$f_{s,min} \leq f_s \leq \min(f_{s,max}, \sqrt{\frac{R_{s,max} r_e}{0.0321}})$$

$$f_{r,min} \leq f_r \leq \min(f_{r,max}, \sqrt{\frac{R_{r,max} r_e}{0.0321}})$$

(20)

Equality Constraint:

- Total depth of cut

The depth of cut in finish pass ($d_s$) and rough passes ($d_r$) are related to the total depth of cut ($d_t$) by:

$$d_t = d_s + n d_r \text{ where } n \text{ is a positive integer} \tag{21}$$

## 3. GA Implementation

The multi-pass milling model used here is based on 7 process variables, speed, feed and depth of cut for finish pass and rough passes and the number of rough passes ($V_s$, $f_s$, $d_s$, $V_r$, $f_r$, $d_r$ and n). Out of these only 6 are independent since $d_s$, $d_r$ and n are related by Eq. 21. The number of variables can be further reduced to 5 by using both the constraints Eq. 14 and Eq. 21. This is done by generating a lookup table which consists of all possible finish and rough depth of cut ($d_s$, $d_r$ respectively) pairs for a given total depth of cut ($d_t$). Since $d_s$ and $d_r$ can take only discrete values from a given set as given by Eq. 14, the number of pairs of ($d_s$,$d_r$) which can be used to remove $d_t$ is limited and is obtained by using Eq. 21. The 5[th] variable then denotes the position of a ($d_s$,$d_r$) pair in this lookup table of feasible ($d_s$,$d_r$) pairs, with $V_s$, $V_r$, $f_s$, $f_r$ as the first 4 variables. This converts the problem into a mixed integer nonlinear problem (MINLP) with four continuous variables and one discrete variable.

Binary coded genetic algorithm with an elitist strategy for replacement is used in this technique. It operates on the principle of the "survival of the fittest" [22, 23]. With MINLP problems, GA generally suffers from the prematurity problem and may require many runs to avoid the trap of local minima [24]. However, the problem being investigated is a special case of the general MINLP problem as the constraints are independent of the integer variable (position in the lookup table). With a suitable fine tuning of the parameters for the genetic operators it is possible to obtain the global optimum results using the proposed technique. A block diagram of the technique is presented in Fig. 2. The technique is described below.

### 3.1. Binary Coding

In binary coded GA, a binary string is used as solution string to represent real values of a variable. The length of the string depends on the precision required [22]. To obtain the actual values of the variables from their binary encoding the following relationship is used:

$$x_i = x_i^{LB} + DV \frac{(x_i^{UB} - x_i^{LB})}{2^{l_i} - 1}$$

*where*

$x_i^{U,LB}$ is upper and lower bound of $i^{th}$ variable

$l_i$ is length of $i^{th}$ string

$DV$ is the Decimal Value of the string

$$DV = \sum_{j=0}^{j=l_i-1} b_j 2^j, \; b_j \text{ is the bit at position } j \text{ of the binary string}$$

(22)

### 3.2. Solution Representation

$V_s$, $V_r$, $f_s$, $f_r$ are the first 4 variables and the 5$^{th}$ variable is the position of the feasible ($d_s$, $d_r$) pair in the lookup table. The first 4 variables are represented by 15 bit binary strings whereas the length of the binary string for 5$^{th}$ variable depends on the table size. All the variable strings are combined to form a single binary string representing the member.

### 3.3. Initialization

The GA is performed using a population size of N=750. Based on the schema theory for binary coded GA, the population size N is chosen by maximizing the rate of gain of the number of schemata represented by binary string of a particular length [22]. In the current study N is chosen to be 750 as it is found that the gain saturates around this value. For initialization 750 different binary strings are created randomly. To create a string, for each bit position a random number is generated in the range [0, 1.0]. The bit is assigned a value '0' if the number turns out to be less than or equal to 0.5 and assigned '1' otherwise.

### 3.4. Evaluation

#### 3.4.1. Fitness Value

To obtain the fitness value corresponding to a member string, first the solution string is broken down into its blocks corresponding to each variable. The variable values are then obtained using the binary coding formula Eq. 22. The decoded value for the first four blocks directly gives the values for $V_s$, $f_s$, $V_r$ and $f_r$ respectively. The depths of cut values are decoded by using the fifth variable (as the position in the lookup table) to pick up the corresponding $(d_s, d_r)$ pair from the lookup table. The objective function is then evaluated using Eq. 11. The objective function value is used directly as the fitness value for the member.

#### 3.4.2. Constraints Handling

To handle constraints, each solution is assigned a constraint violation (CV) value such that all feasible solutions have a zero CV value. For the infeasible solutions it has a non zero value. CV is obtained as:

$$CV = \sum_{j=1}^{J} \langle g_j(\bar{x}) \rangle + \sum_{k=1}^{K} |h_k(\bar{x})|$$

where

$$\langle \alpha \rangle = \begin{cases} -\alpha & \text{if } \alpha < 0 \\ 0 & \text{if } \alpha > 0 \end{cases}$$

Such that the constraints are normalized and in the form:

$$g_j(\bar{x}) \geq 0, \ j = 1, 2, ..., J$$
$$h_k(\bar{x}) = 0, \ k = 1, 2, ..., K$$

(23)

In this technique, only the inequality constraints are used for CV evaluation (Eq. 16, Eq. 18 and Eq. 20) as the equality constraint (Eq. 21) is automatically satisfied when the lookup table for feasible $(d_s, d_r)$ pairs is generated.

### 3.5. Selection

Binary Tournament selection has been used for selection. The selection has been done based upon three possible cases. For the case when both the solutions are feasible, the one with lower function value (i.e. better fitness) is selected. When both the solutions are unfeasible, the one with lower constraint violation is selected. When one of the solutions is feasible and the other infeasible, the feasible solution is selected. For picking up two solutions for comparison, the population is divided into two parts midway and then a solution occupying same position from each part is picked. The population obtained after selection operation is termed as P'.

### 3.6.  Genetic Operators

The crossover operation is performed on the population P' obtained after selection. A two point crossover has been used with a high crossover probability $p_c$=0.8. A two point crossover with the high crossover probability is used as it helps the diversity preservation better than the single point one [23]. After crossover, mutation of the population is performed. Bit wise mutation operator has been used with a mutation probability $p_m$=0.05. In the present study it is found that mutation probability lower than 0.05 leads to premature convergence to a local optimum point due to loss of diversity in the population. Details of the fine tuning tests performed for obtaining the values of genetic parameters are discussed in section 4.1.

### 3.7.  Replacement Strategy: Elitist

Once crossover and mutation is performed over the selected population, the intermediate population is merged with the original population and then the best N out of the 2N members are chosen to entirely replace the original population. This process ensures that the best members of the merged population are not lost and better solutions move into the next generation. Though all members in a population can expect to have a lifetime of one generation, members with a higher fitness can have a longer lifetime when elitist strategies are used [25]. This leads to a higher rate of convergence of the algorithm.

In this work, the best N solutions are identified through the following process:
First, the original population and new populations (obtained after selection, crossover, and mutation) are merged and separated into two distinct portions, a feasible one and an infeasible one. Next, the following cases are considered:

Case 1: Number of feasible solutions > N
The feasible part of the merged population is sorted in increasing order of the objective function value. As we are interested in minimizing the objective function, the top N members are then picked to replace the original population.

Case 2: Number of feasible solutions < N

All the feasible solutions are picked to form the initial population for next generation. But in order to maintain a constant size population some infeasible solutions must also be included. For this the infeasible portion of the merged population is sorted in increasing order of the constraint violation (CV) value. As many of these as required to complete the population are then picked for the next generation.

Case 3: Number of feasible solutions = N

All the feasible solutions are picked to replace the original population.

### 3.8. Convergence Criterion

All the steps from selection to replacement are repeated for a fixed number of generations. The appropriate number of generations is obtained by observing the number of generations for a few trial runs when the difference between the best member and the population average falls below a threshold value. This value is then used for all other future runs. In the current study the GA is run for 100 generations.

## 4. Case Study

In this section, the proposed method has been illustrated in a case study. The data for the example is provided in Table 1, which is same as that used by Shunmugam et al. [14] and An and Chen [15]. Section 4.5 compares the performance of the proposed optimization scheme with the two stage optimization method of Shunmugam et al. and An and Chen. Shunmugam et al. used GA to solve the problem in two stages. In their method, an "unequal depth of cut" strategy is adopted. In the first stage, solutions corresponding to separate minimum costs for the individual rough and finish passes are determined and tabulated for various fixed values of the depth of cut. In the second stage, the optimum number of passes and the optimum subdivision of the depths of cut for different passes are obtained using GA. An and Chen used a similar method for the first stage and an integer programming based method for the second stage.

### 4.1. Selection of GA parameter values

GA is performed with a population size of N=750. The value of N is based on the schema theory for binary coded GA [22]. For binary coded GA, the number of schemata (S) represented by a population of size 'µ' consisting of binary strings of length 'l' is given by:

$$S(\mu,l) = \sum_{i=0}^{l} \binom{l}{i} 2^i \left\{ 1 - \left[1 - \left(\frac{1}{2}\right)^i\right]^{\mu} \right\} \qquad (24)$$

The population size (N) is chosen so that all points in the search space are represented by the initial set of schemata. For this purpose the rate of gain in the number of represented schemata with an increase in population size is maximized. The rate of gain is given by:

$$G = \frac{S(\mu,l) - 2^l}{\mu} \qquad (25)$$

For a binary string length of 65 we calculated G by varying μ in the range [1 5000]. It is found that the maximum value of G is obtained at μ=5000 ($G_{max}$=3.4885e19). Since this value is very high, a practical value of N is chosen corresponding to the point where G=0.999$G_{max}$. This is obtained at μ=750. Hence, a population size (N) of 750 is chosen.

For cross-over probability ($p_c$), a trial value of 0.8 is chosen. With these values of N and $p_c$ and a high value of mutation probability ($p_m$=0.1), GA is run and evolution of the population average fitness and best fitness value is observed. For one value of total depth of cut ($d_t$=6 mm), results are shown in Fig. 3. From this plot it can be seen that the average fitness value approaches the population best as the generation increases. GA is run several times for the same conditions and the generation number corresponding to the generation when the difference between population average and population best falls below 0.25% is observed. It is found to be in the 80-90 range. A conservative value of 100 is hence chosen for the number of generations needed for convergence. It is expected that with lower values of mutation probability, convergence would be faster.

To determine the value of $p_m$, variation of success rate with change in $p_m$ was observed. The success rate here is defined as the percentage of runs for which the GA converged to the global optimum after 100 generations. At each value of $p_m$, 20 GA runs were performed with different initial populations to determine the success rate. From Fig. 4 it can be seen that for $p_m$=0.05 onwards, the success rate is 100%. Hence, $p_m$=0.05 has been selected for solving the problem.

### 4.2. Obtaining look-up table

For a total depth of cut ($d_t$), several combinations of rough and finish cut depths ($d_r$ and $d_s$ respectively) can be obtained which give the same total depth of cut, subject to the constraint Eq. 21. This is done by first making all combinations of ($d_s$,$d_r$) pairs from the range of possible values. Then, for each pair it is checked whether the quantity $\frac{d_t - d_s}{d_r}$ is an integer. The integer value represents the number of rough passes. This condition is equivalent to Eq. 21. A feasible solution is one for which this condition is true. The feasible ($d_s$,$d_r$) pairs for $d_t$=6mm are shown in Table 2. This set of feasible ($d_s$,$d_r$) pairs forms the lookup table discussed in section 3. Each feasible solution is identified by the corresponding pair number in the lookup table. Instead of considering $d_r$ and $d_s$ as two distinct variables, this pair number is considered as the variable for GA. This reduces the number of variables by one. The number of feasible ($d_s$,$d_r$) pairs depends on the value of $d_t$ and for each $d_t$ a new lookup table is generated which satisfies the constraint Eq. 21 for the corresponding $d_t$.

### 4.3. Computation Results

GA was performed on the above problem with different total depths of cut ($d_t$). The algorithms were coded in Visual C++ and run on a Pentium 4 PC. After 100 generations of GA run, the member with lowest objective function value in the population was chosen as the optimum value. The corresponding optimum machining parameters for different total depths of cut are tabulated in Table 3. The tool life for rough and finish passes can be obtained from the optimum machining conditions by using the tool life equation (Eq. 10). The zero values for constraint violation (CV) suggest that all the solutions are feasible and do not violate any of the constraints.

### 4.4. Verification of global optimum

For the problem of depth of cut distribution into multiple passes, if we disregard the optimization criterion of minimizing the unit cost, then we can have several ways of distributing the total depth of cut into 'n' number of rough passes and a single finish pass. All such choices of ($d_s$,$d_r$) pairs are tabulated in the look-up table before starting with the GA optimization. Now, if we select any one of these set of rough and finish depth of cut, we can separately optimize the cutting velocity and feed for the rough and finish passes. The unit cost for finish pass ($UC_s$) and unit cost for rough pass ($UC_r$) can be separately minimized. As the depths of cut are already known, only two variables are involved (cutting speed and feed). Thus for each pair of feasible ($d_r$,$d_s$) pair, a pair of cutting velocity and feed can be obtained for minimum $UC_s$ and $UC_r$. In this way, we can have several optimum solutions for unit cost UC (Eq. 11) corresponding to each feasible set of ($d_r$,$d_s$). However, not all of these unit costs will have the same value. The solution corresponding to lowest unit cost among these is then called as the global optimum and all other solutions are the respective local optimum.

To verify whether the results obtained from GA are the global optimums, the local minimum UC corresponding to each combination of feasible ($d_r$,$d_s$) pair for $d_t$= 6mm is calculated. Optimization of $UC_s$ and $UC_r$ has been done analytically by obtaining the derivates of the functions $UC_s$ and $UC_r$ in terms of the cutting velocity and feed and using the classical conditions for minimization of two variable functions. UC is then calculated by using Eq. 11. The local optimum values so obtained are shown in Table 4. It can be seen that the solution obtained from the GA runs for $d_t$= 6mm shown in Table 3 (UC=1.4108 $/piece) matches closely with the global optimum (UC=1.4102 $/piece) as shown in Table 4.

### *4.5.* Comparison with other schemes

A comparison of the proposed scheme with the results of An and Chen [15] and Shunmugam et al. [14] is made in Table 5. The table shows that the minimum unit production cost obtained in the current study are lower than those reported by An and Chen and Shnumugam et al. For $d_t$=8 mm the GA results are better than those of An and Chen by 5.2% and better by 14% than those of Shunmugam et al. It is to be noted that, while calculating the cutting speeds at each pass, both Shunmugam et al. and An and Chen have used an a priori value of the tool life ($T_r$=$T_s$=240 min.) to reduce the number of variables in the first stage. Because of this assumption about tool life, non-optimum cutting velocities are obtained by them for the cases where the cutting velocity obtained by their method does not violate the power constraint. In our proposed single stage method, the tool life values are obtained a posteriori, after the optimum conditions are obtained. From Fig. 5 it is evident that for the optimum conditions, the tool life for rough passes ($T_r$) is not constant and shows wide variations with changes in total depth of cut ($d_t$). The reason for this is discussed in section 4.7 with reference to the relationship among the cutting parameters for the optimum conditions. The tool life for finish pass shows very small variations with changing $d_t$, however the average value for optimum conditions is found to be 220 min. as opposed to the $T_s$=240 min. value used by Shunmugam et al. It is interesting to note that *for the optimum conditions*, the tool life in rough passes is much higher than the tool life in the finish pass. This can be explained by the higher values of the cutting velocity for finish pass for optimum conditions (Table 3). Due to high feed in the rough passes, the cutting velocities of rough passes are limited by the power constraint (Eq. 18) to lower values.

### 4.6. Post optimality constraint sensitivity analysis

A study was made to observe how the optimal values change when the force and power constraints are changed. Optimal unit production costs were obtained by multiplying a factor to the maximum allowable force ($F_{max}$) and maximum allowable power ($P_{max}$). The results were used to compare the sensitivity of the solution with respect to these constraints. As expected, the optimal costs are lower when the constraints are relaxed, i.e. when the maximum allowable force or power limits are higher. From Fig. 6 it can be observed that the optimal cost is more sensitive to the power constraint. From the slope it can be seen that the rate of change of the optimal cost with changes in $P_{max}$ is higher than that for the rate of change of optimal cost with the changes in $F_{max}$. This information is especially useful in the cases where estimated or approximate values for the maximum force and power limits are used and a range of values for $F_{max}$ and $P_{max}$ are possible. Based on Fig. 6, given a choice of relaxing the limits, the power constraint must be relaxed first in order to obtain lower optimal costs. Such constraint sensitivity analysis can be easily done when a GA based method is used for optimizing the speed and feed along with the depth of cut. It is difficult to do the same using classical methods such as the integer programming method used by An and Chen [15] or the two stage method used by Shunmugam et al. [14].

### 4.7. Analysis of relationship in decision space for optimum conditions

To investigate the relationship among the optimum cutting conditions, optimal cost values and total depth of cut ($d_t$), the optimum cutting conditions and the optimal cost were obtained for different $d_t$. The optimal costs are shown in Fig. 7. A similar stepped form nature can be seen when the number of rough passes (n) for the optimum conditions is plotted versus $d_t$ (Fig. 8). This stepped nature can be explained if the effect of increase in 'n' on optimal costs is more than the effect of increase in the depth of rough cut ($d_r$). This suggests that for optimum conditions, the number of rough passes must be as low as possible subject to the available ($d_s$,$d_r$) combinations for a given $d_t$.

When the depth of cut in rough passes ($d_r$) is plotted against the total depth of cut we again find a stepped form plot (Fig. 9). For the same number of rough passes, the optimal cost increases with the increase in $d_r$.

When the feed rate for the rough pass ($f_r$) is plotted against the total depth of cut we find a stepped plot but with the opposite trend (Fig. 10). For the same 'n', optimal cost increases with decreasing feed rate. This figure explains the nature of the $T_r$ versus $d_t$ plot (Fig. 5) since the tool life increases with decrease in the feed rate. Although the depth of cut ($d_r$) increases, the effect of increased feed rate is more dominant because of the higher exponent of feed rate in extended Taylor's tool life equation (Eq. 10).

### 4.8. Strategy for estimating the optimum conditions

From Table 3 it is evident that *for the optimum conditions*, feed rate for finish pass and cutting velocity for finish and rough passes remain almost unchanged with changes in the total depth of cut ($d_t$). The feed rate for finish pass is restricted to a fixed value because of the surface roughness constraint (Eq. 20). Similarly, the cutting velocities in finish and rough passes are restricted to fixed values by the power constraint (Eq. 18). This information along with the relationships among the decision variables for optimum conditions discovered in Section 4.7 can be used to develop an optimization strategy specific to this problem. This strategy allows us to quickly estimate the optimum conditions when the total depth of cut is changed without running the GA again for the new $d_t$.

To start with, we can easily determine the number of rough passes (n) in the optimum condition by finding out the smallest possible integer value for 'n', given the limits for $d_s$ and $d_r$.

$$n = [\frac{d_t - d_{s,\max}}{d_{r,\max}}], \text{ where [] is the greatest integer operator} \tag{26}$$

From the possible ($d_s$,$d_r$) combinations for the given $d_t$, the one which has 'n' number of rough passes is chosen. From Section 4.7, for the optimal conditions the depth of cut in rough pass should be the lowest one if multiple ($d_s$,$d_r$) combinations are possible for the same 'n'. Thus, we have the parameters 'n', '$d_s$' and '$d_r$'.

Next, we use the average values from Table 3 for the feed rate of finish pass ($f_s$=0.279 mm/tooth), the cutting velocity of finish pass ($V_s$=123.2 m/min) and rough passes ($V_r$=60.35 m/min).

Finally, it is known that the cost of a single pass operation decreases with the increase of the feed rate [6]. Hence for the rough pass feed rate ($f_r$) the maximum possible value is chosen subject to constraints given by Eq. 16 and Eq. 20, i.e.

$$f_r = \min\{f_{r,\max}, \sqrt{(R_{r,\max} r_e)/0.0321}, f_r^*\} \tag{27}$$

Where $f_r^*$ is obtained from the upper limit of the constraint equation (Eq. 16).

The above procedure is used for estimating the optimum cost for $d_t$=11.5 mm. The optimum values obtained by this strategy and by the GA run are tabulated in Table 6. It can be seen that the cost obtained by this strategy (UC=2.2194 $/piece) is in close agreement with the cost obtained by GA (UC= 2.1995 $/piece). The value obtained by the estimation strategy is poorer from the value obtained from GA by not more than 1%.

This optimization strategy can be used to quickly obtain an estimate of the optimum cutting conditions when the total depth of cut is changed. However, it is to be noted that this strategy is applicable only to the specific problem under consideration and does not work if the parameters of the problem such as number of teeth of cutter or length of work piece are changed. If such a strategy is required for the new problem, then the relationships among the optimum conditions must be re-evaluated after running the GA based method for different $d_t$ values.

## 5. Conclusion

In this paper an optimization methodology is proposed for the optimization of the multi-pass face milling process. Binary coded genetic algorithm (GA) is used to minimize the unit production cost along with the satisfaction of several nonlinear constraints. Equal depth of cut is used for the roughing passes and the relationship between total depth of cut ($d_t$), depth of cut for finish pass ($d_s$) and for rough pass ($d_r$) is represented by a single variable denoting the position of the ($d_s$,$d_r$) pair in the lookup table consisting of all feasible ($d_s$,$d_r$) pairs. This transforms the problem into a mixed integer nonlinear optimization problem. An elitist binary coded GA is used to solve the problem and the entire technique is demonstrated in a case study. The minimum unit production costs obtained by the current technique are better than those reported in literature for the same problem [14, 15]. The numerical test shows that GA does not get stuck to the local optimums and is successful in converging to the global optimum. On performing post optimality analysis for constraint sensitivity it is found that the optimum point is more sensitive to the power constraint as compared to the force constraint. Based on the analysis of the optimum results, relationships among the decision variables are identified. These relationships are used to develop a strategy to quickly estimate the optimum cutting conditions without running the GA, when the total depth of cut ($d_t$) is changed. The prediction of the estimation strategy is found to match closely to the results obtained from GA.

## 6. Scope for future development

In the present work, all rough passes have been considered to have the same depth of cut. However, it still needs to be investigated whether a strategy of unequal depth of cut for rough passes can provide better results using this technique. It would be an interesting problem to identify the conditions under which an unequal depth of cut would be better than the equal depth of cut strategy and vice versa.

The main focus of the present paper is on showing the effectiveness of the proposed optimization methodology. Therefore, simple relations have been used for the estimation of cutting forces and surface roughness. However, even with more realistic relations the same optimization technique can be used. Additional constraints may also be easily introduced to make the optimization problem more realistic. The lookup table based technique employed in this paper can be extended for optimization of other multi-pass machining processes.

# Nomenclature

| | |
|---|---|
| $B$ | Width of work piece (mm) |
| $C_0$ | Constant in tool life equation for face milling |
| $C_1, C_2$ | Constants in cutting force and power equations respectively |
| $D$ | Cutter diameter (mm) |
| $d_r$ | Depth of cut for the rough passes (mm) |
| $d_s$ | Depth of cut for finish pass (mm) |
| $d_t$ | Total depth of cut (mm) |
| $e_r, e_s$ | Extra travel of cutter at the ends (mm) |
| $F$ | Cutting force (kgf) |
| $F_{max}$ | Maximum allowable cutting force (kgf) |
| $f_r$ | Feed for the rough passes (mm/tooth) |
| $f_s$ | Feed for the finish pass (mm/tooth) |
| $h_1$ | Constant related to tool travel time (min/mm) |
| $h_2$ | Constant related to tool approach/depart time (min) |
| $k_o$ | Labor cost per unit time including overhead cost ($/min) |
| $k_t$ | Cost of cutting tool per edge ($/cutting edge) |
| $L$ | Length of work piece (mm) |
| $L_{tr}$ | Cutting travel length for rough pass (mm) |
| $L_{ts}$ | Cutting travel length for finish pass (mm) |
| $N$ | Population size for GA runs |
| $n$ | Number of rough passes |
| $n_1, n_2, n_3$ | Exponent of cutting velocity, depth of cut and feed in tool life equation |
| $n_4, n_5$ | Exponent of depth of cut and feed in the empirical relation for cutting force |
| $P$ | Cutting power (kW) |
| $P_{max}$ | Maximum power (kW) |
| $p_c, p_m$ | Crossover and mutation probability in GA runs |
| $R_{r,max}, R_{s,max}$ | Maximum allowable surface roughness (Ra value) for rough and finish passes (mm) |
| $r_e$ | Cutter nose radius (mm) |
| $T_r, T_s$ | Tool life for rough and finish passes (min) |
| $t_e$ | Tool exchange time (min/cutting edge) |
| $t_i$ | Idle tool motion time (min) |
| $t_l$ | Machine idling time (min) |
| $t_{mr}, t_{ms}$ | Actual machining time for rough and finish passes (min) |
| $t_p$ | Preparation time (min/piece) |
| UC | Total machining cost per unit ($/piece) |
| $UC_r, UC_s$ | Machining cost per unit for rough and finish passes ($/piece) |
| $V_r$ | Cutting velocity for rough passes (m/min) |
| $V_s$ | Cutting velocity for finish pass (m/min) |
| $Z$ | Number of teeth in the cutter |

# References

1. Taylor FW (1906) On the art of cutting metals. J Eng Ind Trans ASME 28:31-350

2. Gilbert WW (1950) Economics of machining. Machining theory and practice. American Society of Metals, Cleveland, Ohio, pp 465–485

3. Petropoulos PG (1973) Optimal selection of machining rate variables by geometric programming. Int J Prod Res 11(4):305–314

4. Abuelnaga AM, El-Dardiry EA (1984) Optimization methods for metal cutting. Int J Mach Tool Des Res 24(1):11–18

5. Ermer DS (1971) Optimization of the constrained machining economics problem by geometric programming. J Eng Ind 93:1067–1072

6. Shin YC, Joo YS (1992) Optimization of machining conditions with practical constraints. Int J Prod Res 30(12):2907–2919

7. Al-Ahmari AMA (2001) Mathematical model for determining machining parameters in multipass turning operations with constrains. Int J Prod Res 39:3367-3376

8. Sankar RS, Asokan P, Saravanan R, Kumanan S, Prabhaharan G (2007) Selection of machining parameters for constrained machining problem using evolutionary computation. Int J Adv Manu Tech 32 (9-10): 892-901

9. Lambert BK, Walvekar AG (1978) Optimization of multipass machining operations. Int J Prod Res 16(4):259–265

10. Yellowley I, Gunn EA (1989) Optimal subdivision of cut in multi-pass machining operations. Int J Prod Res 27(9):1573-1588

11. Chen MC, Tsai DM (1996) A simulated annealing approach for optimization of multi-pass turning operations. Int J Prod Res 34(10):2803-2825

12. Sönmez AI, Baykasoğlu A, Dereli T, Filiz IH (1999) Dynamic optimization of multipass milling operations via geometric programming. Int J Mach Tool Manufact 39(2):297-320

13. Gupta R, Batra JL, Lal GK (1995) Determination of optimal subdivision of depth of cut in multipass turning with constraints. Int J Prod Res 33(9):2555–2565

14. Shunmugam MS, Reddy SVB, Narendran TT (2000) Selection of optimal conditions in multi-pass face-milling using genetic algorithm. Int J Mach Tools Manuf 40:401–414

15. An L, Chen M (2003) On Optimization of Machining Parameters. Control and Automation ICCA '03. Proceedings. 4th International Conference on pp.839-843, 10-12 June 2003

16. Jawahir, I.S. and Wang, X (2007) Development of hybrid predictive models and optimization techniques for machining operations, J Mater Process Tech 185:46-59

17. Wang X, Kardekar A, Jawahir IS, Performance-based optimization of multi-pass face-milling operations using genetic algorithms. Intelligent Computation in Manufacturing Engineering Conference, Naples, Italy, June 2004.

18. Sheta A, Turabieh H, Vasant, P (2007) Hybrid Optimization Genetic Algorithms (HOGA) with Interactive Evolution to Solve Constraint Optimization Problems for Production systems. Int J Computational Sc Vol. 1, No. 4:395-406.

19. Wang ZG, Rahman M, Wong YS, Sun J (2005) Optimization of multi-pass milling using parallel genetic algorithm and parallel genetic simulated annealing. Int J Mach Tools Manuf 45(15):1726-1734

20. Nefedov N, Osipov K (1987) Typical Examples and Problems in Metal Cutting and Tool Design. Mir Publishers, Moscow

21. Boothroyd G (1985) Fundamentals of Metal Machining and Machine Tools. McGraw-Hill, New Delhi

22. Holland JH (1975) Adaptation in Natural and Artificial Systems. University of Michigan Press, Second edition: MIT Press, 1992

23. Mitchell M (1998) An Introduction to Genetic Algorithms. MIT Press, Cambridge, MA

24. Yokota T, Gen M, Li YX (1996) Genetic algorithm for non-linear mixed integer programming problems and its applications. Computers and Industrial Engineering 30(4):905-917

25. DeJong K (1975) An analysis of the behavior of a class of genetic adaptive systems. Ph.D. thesis, University of Michigan, USA

**Figure Captions**

Fig. 1: (a) Cutting process in face milling showing the cutter path in (b) rough pass and (c) finish pass

Fig. 2: Block diagram of the proposed genetic algorithm

Fig 3: Evolution of population best and population average value of unit cost (UC) with generation

Fig 4: Variation of success rate of convergence of GA with mutation probability

Fig. 5: Variation of tool life in rough and finish passes for optimum cutting conditions with changes in the total depth of cut ($d_t$)

Fig. 6: Effect of changing the force and power constraints on the optimum values of unit production cost (at $d_t$=6mm)

Fig. 7: Variation of optimum unit production cost with total depth of cut ($d_t$)

Fig. 8: Variation of number of rough passes for optimum conditions with changes in the total depth of cut ($d_t$)

Fig. 9: Variation of depth of cut of rough passes for optimum conditions with changes in the total depth of cut ($d_t$)

Fig. 10: Variation of the feed for rough passes for optimum conditions with changes in the total depth of cut ($d_t$)

## Table Captions

Table 1: Data used for the example problem

Table 2: Look-up table consisting of the feasible combinations of finish and rough depth of cut ($d_s$,$d_r$) pairs for a total depth of cut ($d_t$) of 6 mm

Table 3: Cutting parameter values, total machining cost and tool life corresponding to optimum operation point for different total depths of cut

Table 4: Local optimum unit costs corresponding to all the feasible combinations of finish and rough depths of cut (feasible (ds,dr) pairs ) for a total depth of cut of 6 mm

Table 5: Comparison of the results obtained from the proposed GA based technique with the results of Shunmugam et al. [14] and An and Chen [15]

Table 5: Comparison of the optimization results obtained from GA and the estimation strategy developed in Sec. 4.8 for the total depth of cut ($d_t$) of 11.5mm

# Figures

Figure 1

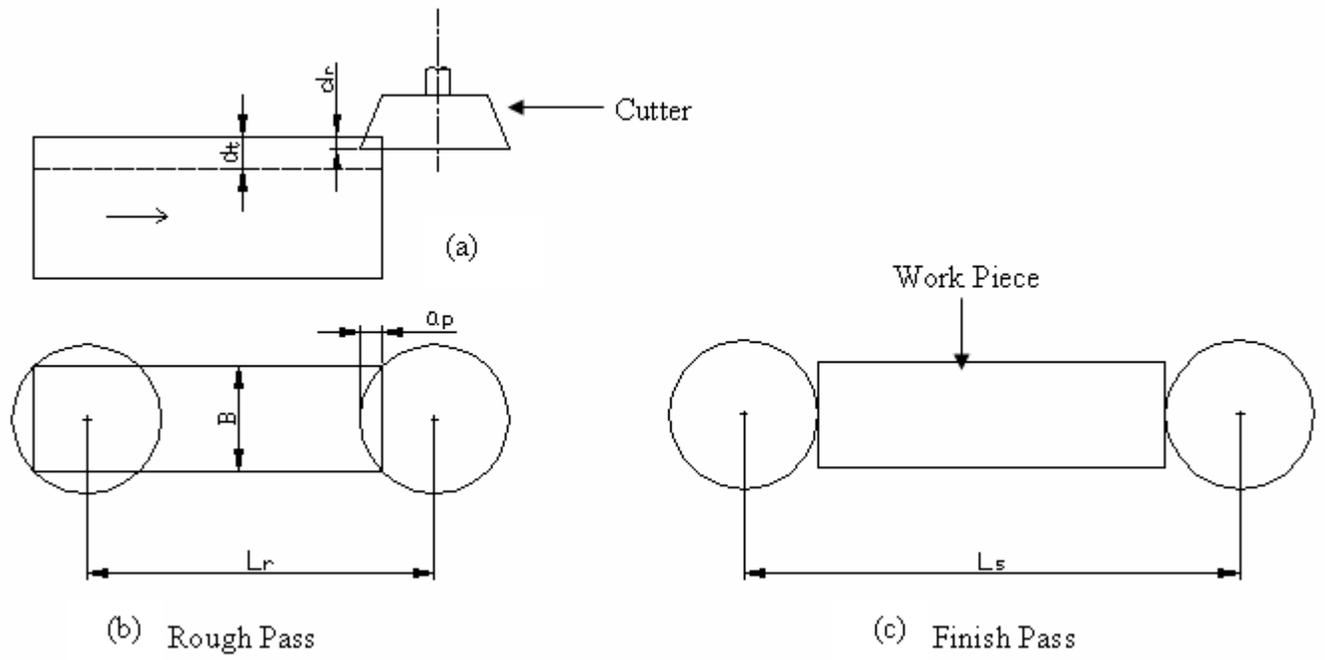

Figure 2

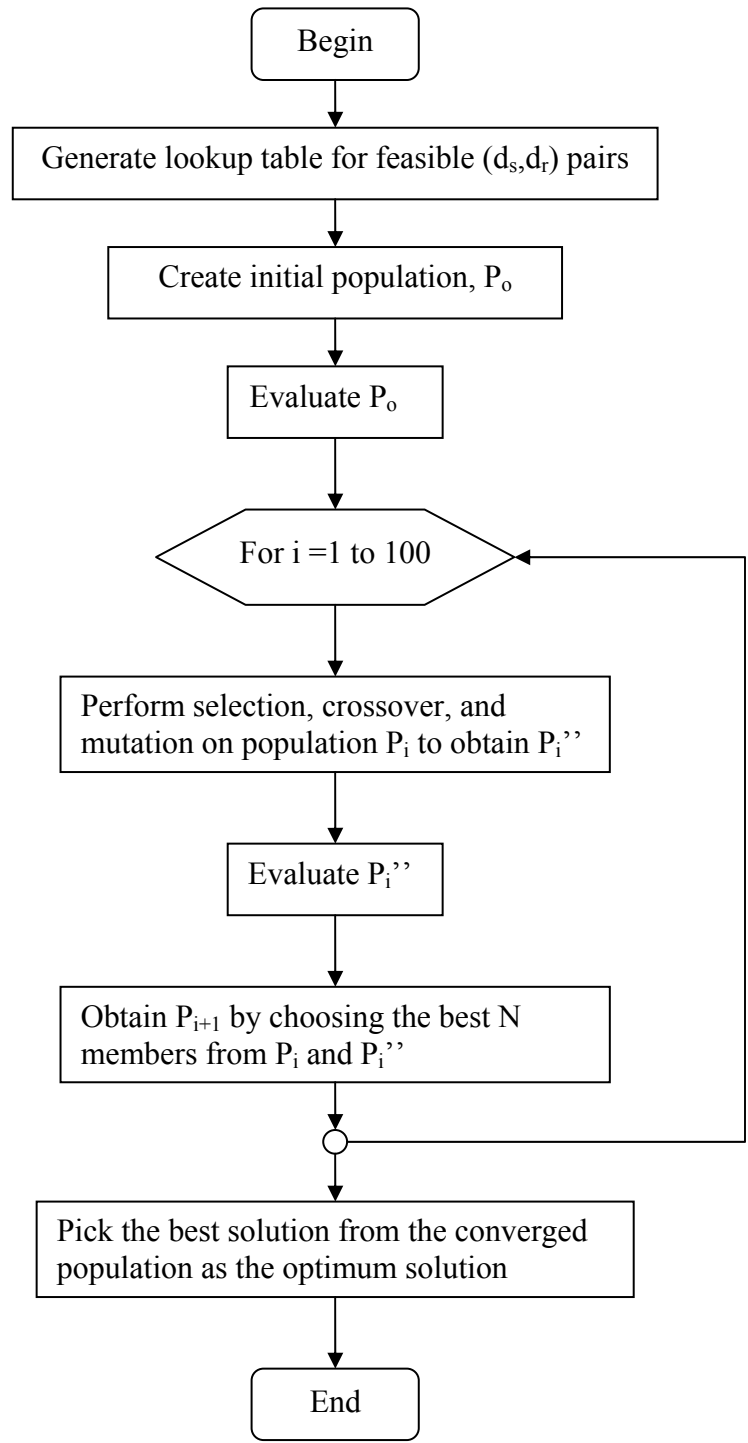

Figure 3

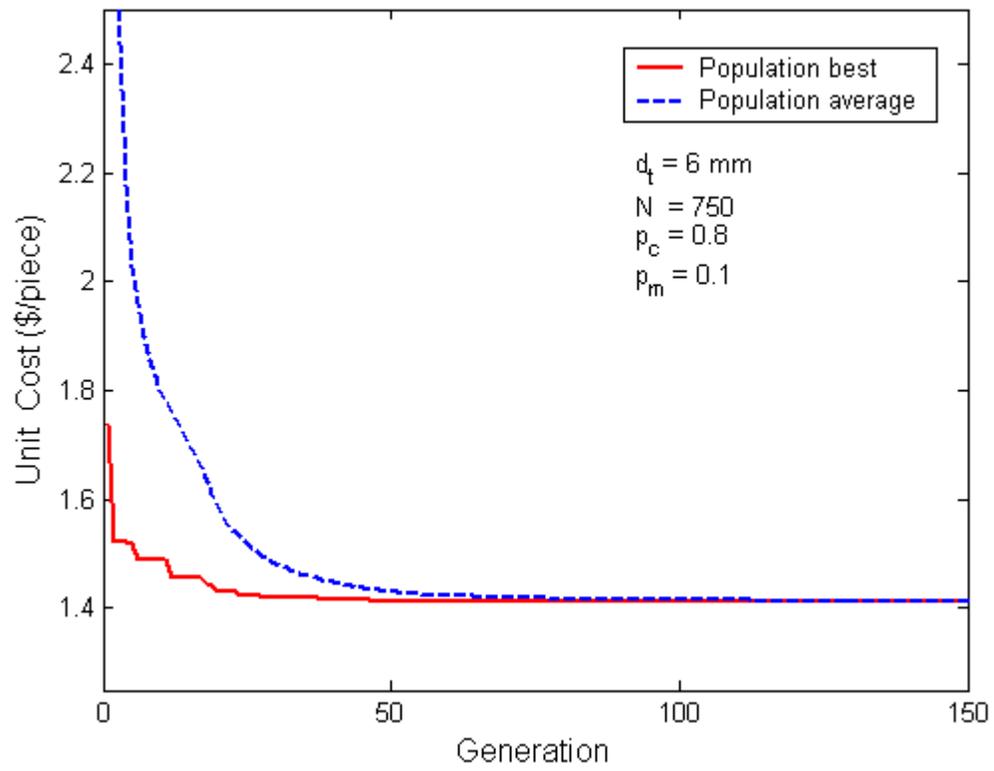

Figure 4

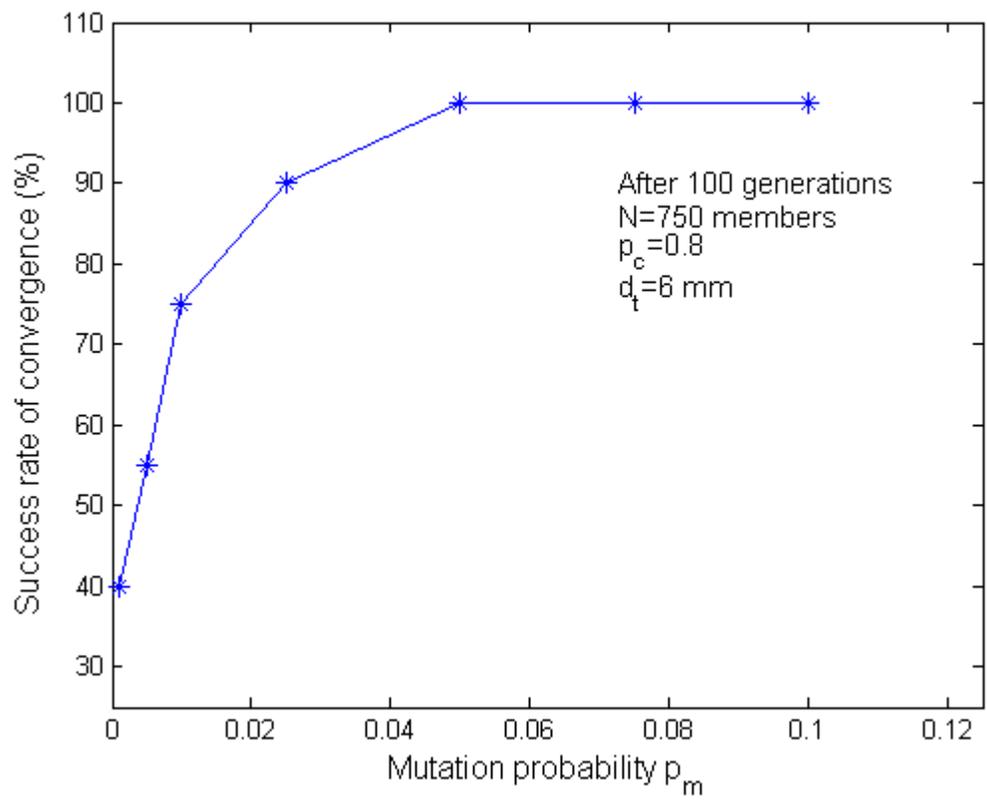

Figure 5

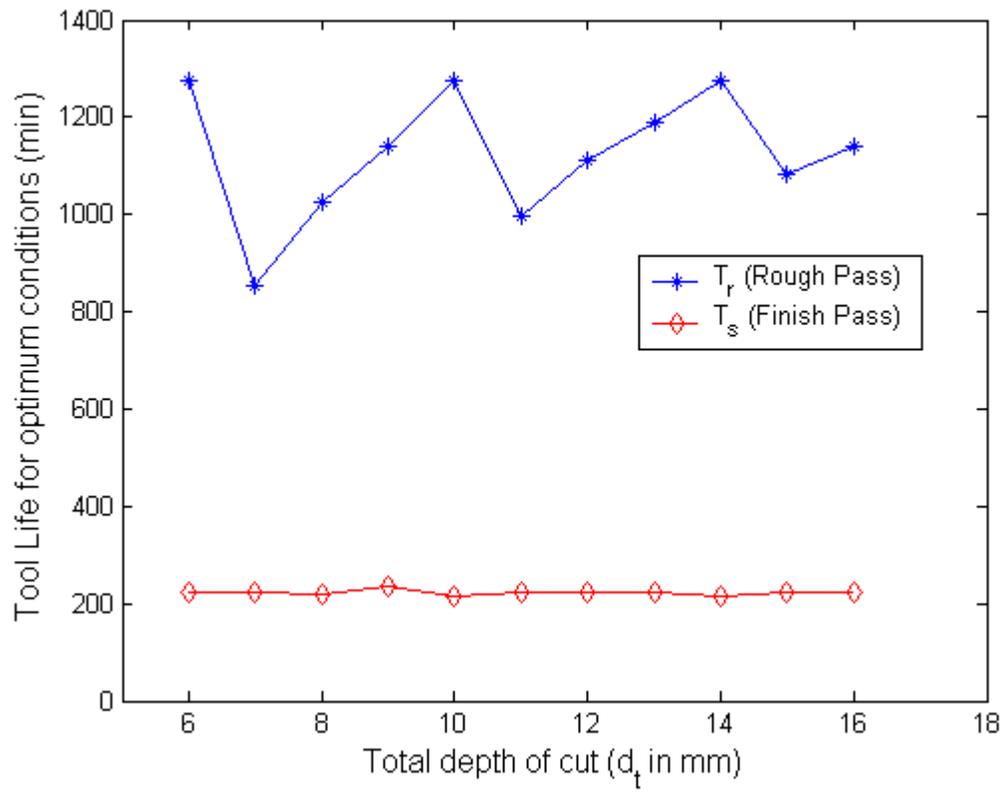

Figure 6

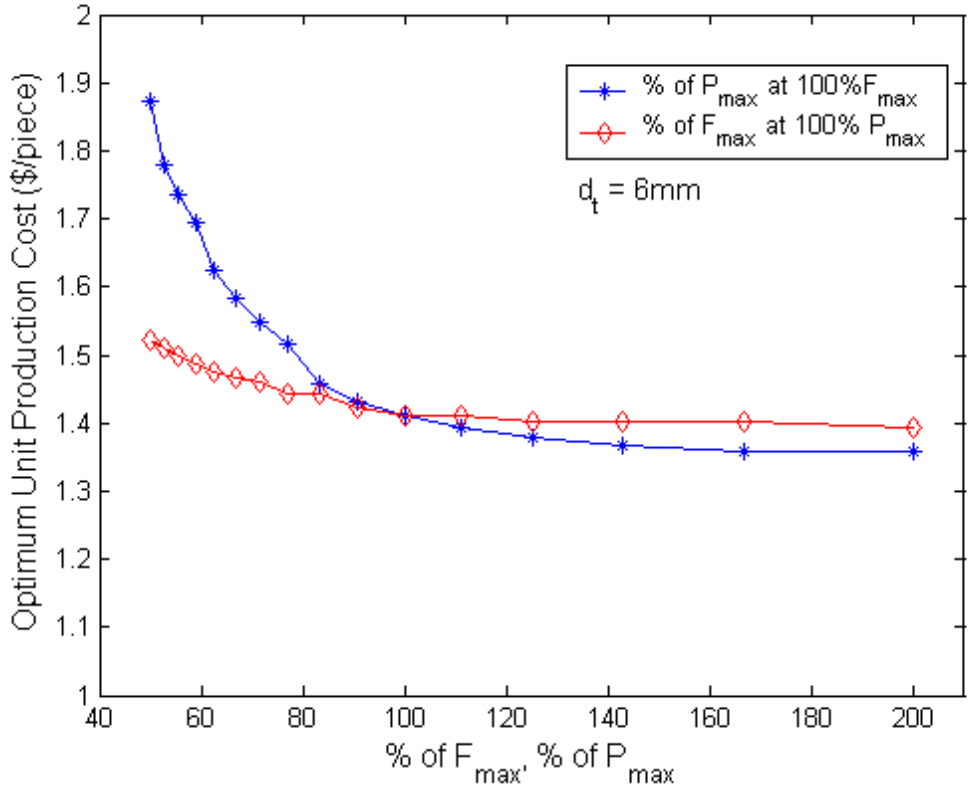

Figure 7

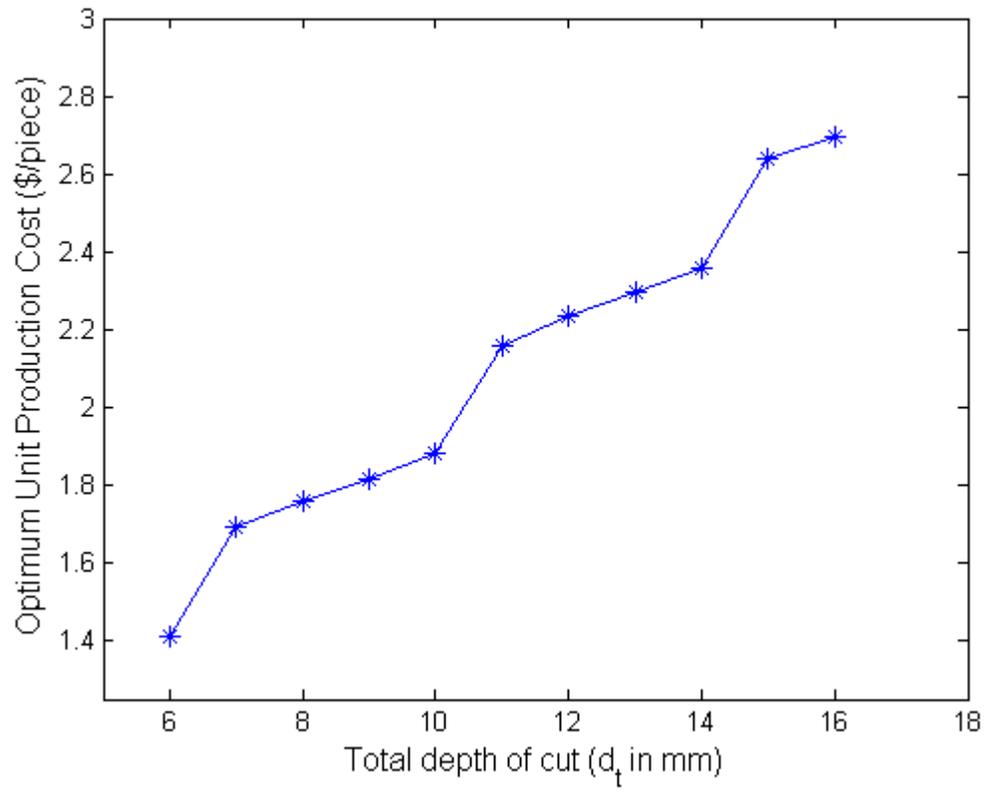

Figure 8

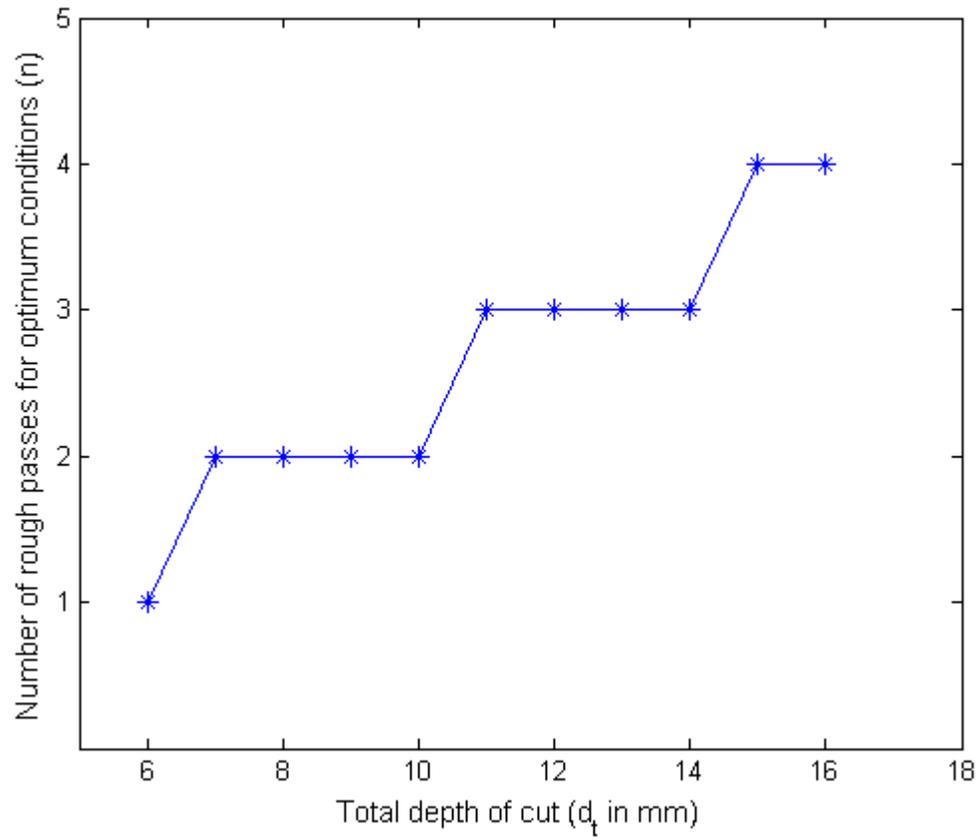

Figure 9

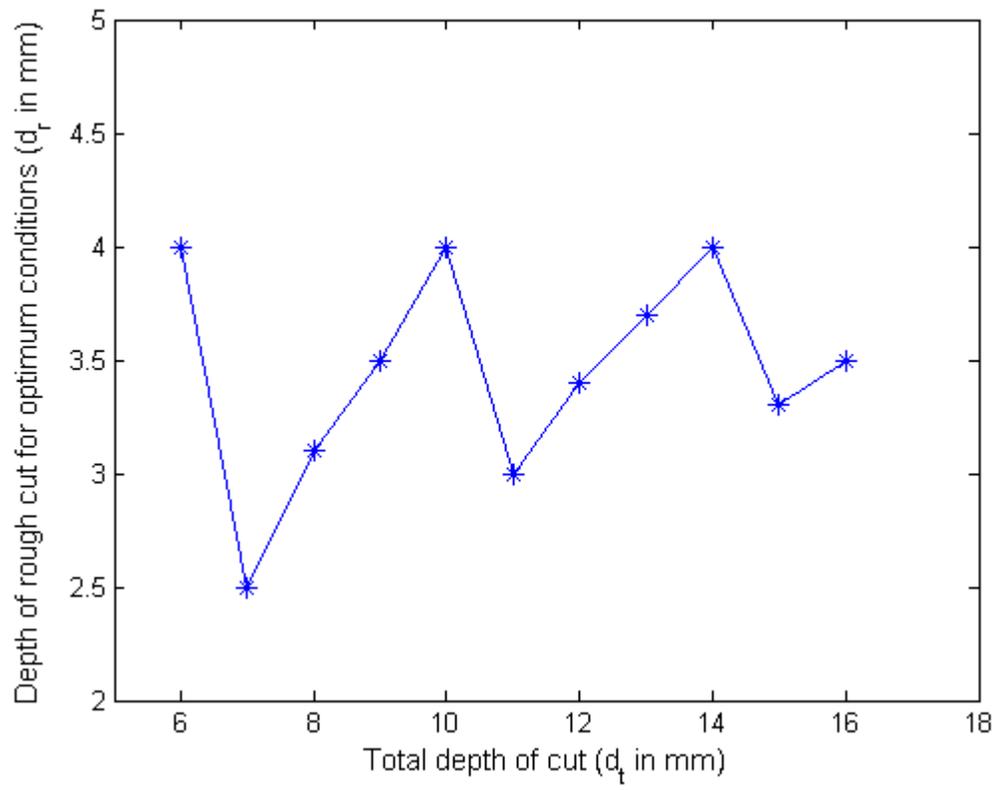

Figure 10

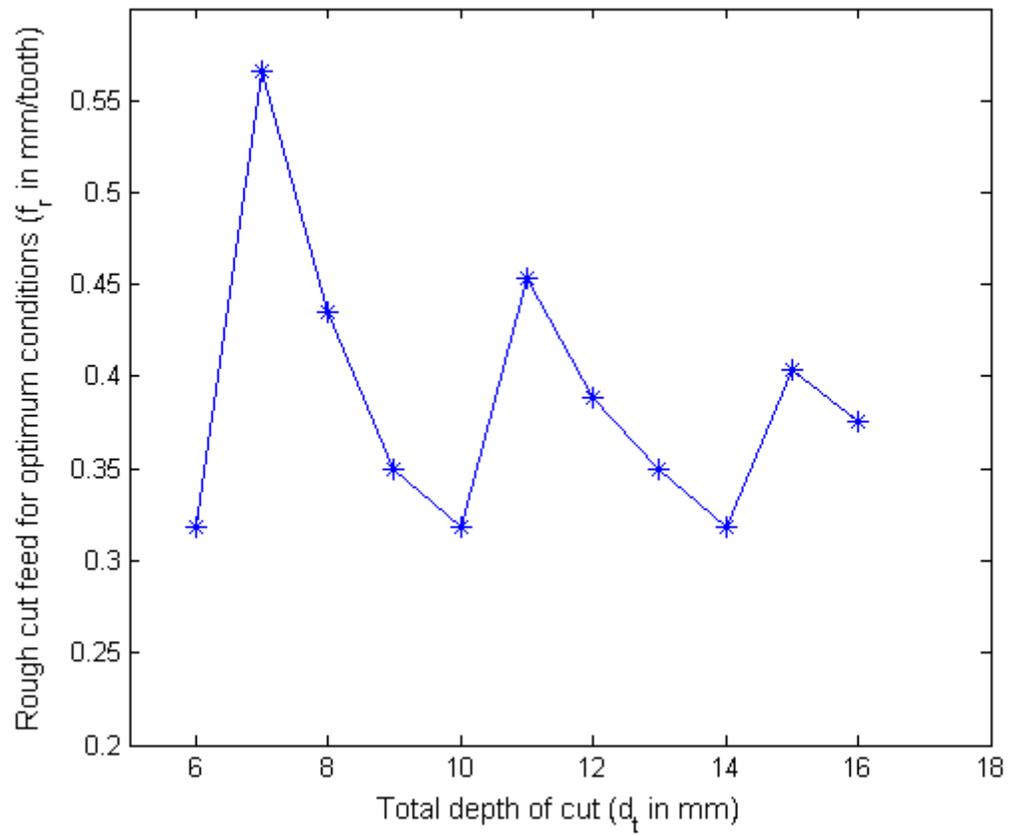

**Tables:**

Table 1

| |
|---|
| L= 400 mm, $L_{ts}$=403 mm, $L_{tr}$=260.55 mm, B=100 mm, D=160 mm, Z=16, $r_e$=1.0 mm |
| $e_r$=$e_s$=3 mm, $h_1$=7x10$^{-4}$ min/mm, $h_2$=0.3 min |
| $k_o$=0.5 \$/min, $k_t$=2.5 \$/cutting edge, $t_e$=1.5 min/cutting edge $t_p$=0.75 min/piece |
| $V_{s,min}$=$V_{r,min}$=50 m/min, $V_{s,max}$=$V_{r,max}$=300 m/min, $f_{s,min}$=$f_{r,min}$=0.1 mm/tooth, $f_{s,max}$=$f_{r,max}$=0.6 mm/tooth, $d_{s,min}$=0.5 mm, $d_{s,max}$=2 mm, $d_{r,min}$=1 mm, $d_{r,max}$=4 mm, $d_{s,step}$=$d_{r,step}$=0.1 mm |
| $F_{max}$=815.77 kgf, $P_{max}$=10 kW, $R_{s,max}$=0.0025 mm, $R_{r,max}$=0.025 mm, η=0.8, |
| $C_v$=445, l=0.32, $x_v$=0.15, $y_v$=0.35, $p_v$=0, $q_v$=0.2, $s_v$=0.2, $K_v$=1.0, $C_f$=54.5, $s_f$=1.0, $p_f$=1.0, $q_f$=1.0, $K_f$=1.0 |
| $C_0$=253337816.7, $C_1$=545, $C_2$=0.111315, $n_1$=3.125, $n_2$=0.46875, $n_3$=1.09375, $n_4$=0.9, $n_5$=0.74 |
| $a_s$=6.330309, $a_r$=4.09271 $b_s$=1.680135x10$^{-6}$, $b_r$=2.598712x10$^{-6}$, $c_s$=0.29105, $c_r$=0.2411925 |

Table 2

| Pair number | $d_s$ mm | $d_r$ mm |
|---|---|---|
| 1 | 1 | 0.5 |
| 2 | 1 | 1 |
| 3 | 1 | 2.5 |
| 4 | 1.5 | 0.5 |
| 5 | 1.5 | 1.5 |
| 6 | 2 | 0.5 |
| 7 | 2 | 1 |
| 8 | 2 | 2 |
| 9 | 2 | 4 |

Table 3

| $d_t$ | $V_s$ | $V_r$ | $f_s$ | $f_r$ | $d_s$ | $d_r$ | n | CV | UC | $T_s$ | $T_r$ |
|---|---|---|---|---|---|---|---|---|---|---|---|
| mm | m/min | | mm/tooth | | mm | | | | $/piece | min | |
| 6 | 122.23 | 60.12 | 0.2791 | 0.3187 | 2 | 4 | 1 | 0 | 1.4108 | 222 | 1274 |
| 7 | 122.38 | 60.00 | 0.2791 | 0.5658 | 2 | 2.5 | 2 | 0 | 1.6914 | 221 | 853 |
| 8 | 124.46 | 60.03 | 0.2790 | 0.4355 | 1.8 | 3.1 | 2 | 0 | 1.7615 | 220 | 1025 |
| 9 | 123.80 | 63.27 | 0.2790 | 0.3499 | 2 | 3.5 | 2 | 0 | 1.8276 | 213 | 1044 |
| 10 | 123.40 | 60.13 | 0.2791 | 0.3187 | 2 | 4 | 2 | 0 | 1.8830 | 216 | 1274 |
| 11 | 122.10 | 60.03 | 0.2790 | 0.4533 | 2 | 3 | 3 | 0 | 2.1606 | 223 | 997 |
| 12 | 124.00 | 60.05 | 0.2789 | 0.3890 | 1.8 | 3.4 | 3 | 0 | 2.2328 | 223 | 1110 |
| 13 | 122.92 | 60.19 | 0.2791 | 0.3499 | 1.9 | 3.7 | 3 | 0 | 2.2940 | 223 | 1189 |
| 14 | 123.70 | 60.13 | 0.2789 | 0.3187 | 2 | 4 | 3 | 0 | 2.3553 | 214 | 1274 |
| 15 | 123.89 | 60.02 | 0.2791 | 0.4037 | 1.8 | 3.3 | 4 | 0 | 2.6396 | 224 | 1082 |
| 16 | 122.07 | 60.03 | 0.2790 | 0.3757 | 2 | 3.5 | 4 | 0 | 2.6956 | 223 | 1138 |

Table 4

| $d_s$ | $d_r$ | n | $UC_s$ | $UC_r$ | UC |
|---|---|---|---|---|---|
| mm | mm |  | $/piece | $/piece | $/piece |
|  |  |  |  |  |  |
| 1 | 0.5 | 10 | 0.53660 | 0.40794 | 4.9910 |
| 1 | 1 | 5 | 0.53660 | 0.34594 | 2.6413 |
| 1 | 2.5 | 2 | 0.53660 | 0.38123 | 1.6741 |
| 1.5 | 0.5 | 9 | 0.55920 | 0.40794 | 4.6057 |
| 1.5 | 1.5 | 3 | 0.55920 | 0.34733 | 1.9762 |
| 2 | 0.5 | 8 | 0.57098 | 0.40794 | 4.2095 |
| 2 | 1 | 4 | 0.57098 | 0.34594 | 2.3297 |
| 2 | 2 | 2 | 0.57098 | 0.36078 | 1.6675 |
| 2 | 4 | 1 | 0.57098 | 0.46420 | 1.4102 |

Table 5

| Total depth of cut ($d_t$ in mm) | Optimum unit production cost, UC ($/piece) | | |
|---|---|---|---|
| | Current Study | An and Chen [15] | Shunmugam et al.[14] |
| 6 | 1.4108 | 1.4858 | --- |
| 8 | 1.7615 | 1.8523 | 2.0086 |

Table 6

| Method | $d_t$ | $V_s$ | $V_r$ | $f_s$ | $f_r$ | $d_s$ | $d_r$ | n | UC,min |
|---|---|---|---|---|---|---|---|---|---|
| | mm | m/min | | mm/tooth | | mm | | | $/piece |
| GA | 11.5 | 123.24 | 60.73 | 0.2791 | 0.4125 | 1.9 | 3.2 | 3 | 2.1995 |
| Estimation Strategy (Sec 4.8) | 11.5 | 123.2 | 60.35 | 0.279 | 0.424 | 1.9 | 3.2 | 3 | 2.2194 |